\documentclass[10pt,twocolumn,letterpaper]{article}

\usepackage{iccv}              

\definecolor{iccvblue}{rgb}{0.21,0.49,0.74}
\usepackage[pagebackref,breaklinks,colorlinks,allcolors=iccvblue]{hyperref}
\usepackage{graphicx,verbatim}
\usepackage{amsmath}
\usepackage{booktabs}
\usepackage{amssymb}
\usepackage{textcomp}
\usepackage{booktabs}
\usepackage{float}
\usepackage{adjustbox}
\usepackage{fancyhdr}
\usepackage{stfloats}
\usepackage{natbib}
\usepackage[T1]{fontenc}
\usepackage[utf8]{inputenc}

\def\paperID{*****} 
\def\confName{ICCV}
\def\confYear{2025}

\title{$\mathtt{MedVisionLlama}$: Leveraging Pre-Trained Large Language Model Layers to Enhance Medical Image Segmentation}

\author{
Gurucharan Marthi Krishna Kumar\\
Montreal Neurological Institute, McGill University\\
{\tt\small gurucharan.marthikrishnakumar@mail.mcgill.ca}
\and
Aman Chadha\thanks{Work done outside role at Amazon.}\\
Amazon\\
{\tt\small aman@amanchadha.com}
\and
Janine Mendola\\
Dept. of Ophthalmology, McGill University\\
{\tt\small janine.mendola@mcgill.ca}
\and
Amir Shmuel\\
McConnell Brain Imaging Centre, \\
Montreal Neurological Institute, McGill University\\
{\tt\small amir.shmuel@mcgill.ca}
}

\begin{document}
\maketitle
\begin{abstract}
Medical image segmentation plays a key role in healthcare, enabling accurate diagnosis and treatment planning. Vision Transformers (ViTs) show strong potential for segmentation tasks, but their dependence on large datasets limits practical usage in clinical settings. This study explores whether integrating pre-trained Large Language Models (LLMs) with ViT-based segmentation models can enhance feature refinement and improve performance in data-constrained environments. We introduce $\mathtt{MedVisionLlama}$, which combines ViT encoders with pre-trained Llama weights and applies Low-Rank Adaptation (LoRA) for fine-tuning in 3D medical image segmentation. Evaluated on the Medical Segmentation Decathlon dataset, the model consistently outperformed a standard ViT, showing improved generalization across MRI and CT modalities. It maintained stable segmentation quality even with limited training data and across varied anatomical structures. Activation maps revealed sharper and more stable attention to relevant regions. Ablation studies confirmed that the performance gains stemmed from LLM-based feature refinement rather than increased model complexity. $\mathtt{MedVisionLlama}$ offers a scalable and data-efficient solution for medical image segmentation. Source code and implementation are available at: \href{https://github.com/AS-Lab/Marthi-et-al-2025-MedVisionLlama-Pre-Trained-LLM-Layers-to-Enhance-Medical-Image-Segmentation}{https://github.com/AS-Lab/Marthi-et-al-2025-MedVisionLlama-Pre-Trained-LLM-Layers-to-Enhance-Medical-Image-Segmentation}.
\end{abstract}
\section{Introduction}
\label{sec:intro}

Medical imaging techniques such as MRI, CT, and X-rays offer non-invasive imaging but suffer from noise, low resolution, and variability which can lead to inaccurate segmentation and misdiagnosis \cite{zhao2024foundation}. Deep learning, particularly Convolutional Neural Networks (CNNs) and Vision Transformers (ViTs) \cite{dosovitskiy2020image,ma2024segment}, has improved segmentation accuracy and efficiency, though consistency across clinical workflows remains a concern. CNNs such as U-Net \cite{ronneberger2015u} effectively capture local features but struggle with long-range context due to their localized operations \cite{chen2021transunet,valanarasu2021medical}. ViTs address this limitation by leveraging global attention \cite{vaswani2017attention}. This complementarity has inspired hybrid CNN-Transformer models such as UNETR \cite{hatamizadeh2022unetr} and Swin-UNETR \cite{cao2022swin}. However, ViTs require large labeled datasets, high computational cost, and careful tuning \cite{lai2024residual}, limiting their use in low-data scenarios.

Large language models (LLMs) show strong generalization, especially in few-shot segmentation \cite{hossain2024visual,zhu2024llafs}. Therefore, they are promising models for handling limited data while maintaining performance across diverse clinical conditions. Prior work on Vision-Language Models (VLMs) has explored textual guidance \cite{yun2023ifseg,li2023lvit} and frozen transformer blocks \cite{pang2023frozen,chen2025multi,man2024situational} for scalable integration and improved feature learning. Building on advances in both classification \cite{lai2024residual} and segmentation \cite{kumari2024leveraging,wang2024adapting}, we explore LLM integration to enhance feature refinement and robustness in low-data ViT-based segmentation tasks.

We hypothesize that pre-trained LLMs can act as residual attention boosters in segmentation, enhancing focus on critical image regions. The \textit{Information Filtering Hypothesis} \citep{pang2023frozen} suggests that LLM transformer blocks filter and enhance key visual tokens. We extend this idea to medical image segmentation, a domain where it remains underexplored. We propose integrating a frozen LLM layer within a ViT-based segmentation model to learn representations at various levels and capture the overall image structure, without requiring extensive task-specific fine-tuning, thereby enabling effective reuse of learned representations.

To validate this approach, we introduce $\mathtt{MedVisionLlama}$, a ViT-based segmentation model enhanced with a pre-trained LLM using Parameter-Efficient Fine-Tuning via Low-Rank Adaptation (LoRA) \citep{hu2022lora}. Our approach uses the frozen pre-trained LLM weights as a visual encoder without prompts, leveraging its learned semantic representations to improve segmentation accuracy. We show improvements in segmentation accuracy, feature refinement, and data efficiency across ten medical imaging tasks, driven by LLM-based feature enhancement rather than added parameter count. This results in a more stable and reproducible model that can be used in settings with limited resources.

By integrating LLMs into ViT-based segmentation models, our method tackles data scarcity and improves performance. Our work highlights a new approach for applying LLMs to address challenges in medical image segmentation. In summary, our paper presents the following key contributions:

\begin{list}{$\bullet$}{}
    \item We show that frozen pre-trained LLM weights with LoRA improve segmentation accuracy by refining attention mechanisms, validated through activation map analysis across diverse anatomical structures.
    \item We demonstrate that the proposed model outperforms a standard ViT in few-shot settings, generalizing well from limited training data across medical imaging tasks.
    \item We establish that performance improvements result from LLM-driven feature refinement rather than increased model parameters, with general and medical LLMs providing comparable benefits despite additional complexity.
\end{list}

\section{Related Studies}
\label{sec:related}

\subsection{Vision Transformer for Medical Image Segmentation}

While CNN-based methods such as U-Net \cite{ronneberger2015u} have driven progress in medical segmentation by efficiently capturing local features, their limited ability to model long-range dependencies remains a challenge \cite{chen2021transunet,valanarasu2021medical}. Transformers \cite{vaswani2017attention}, initially developed for textual data, overcome this limitation through self-attention, which captures global spatial relationships. ViTs demonstrated the potential of pure self-attention models in vision tasks. Subsequent models such as Swin Transformer \cite{liu2021swin}, PVT \cite{wang2021pyramid}, and CvT \cite{wu2021cvt} further optimized transformer-based segmentation. Hybrid CNN-Transformer models have gained popularity in medical imaging \cite{hatamizadeh2022unetr,huang2021missformer,cao2022swin,heidari2023hiformer,kolahi2024msa}, combining local and global modeling for improved segmentation across varied imaging protocols.

\subsection{Large Language Models}

Although originally developed for natural language processing, LLMs have recently shown promise in vision and medical imaging tasks by offering strong generalization, contextual reasoning, and modular integration with visual encoders. Pre-training on large datasets enables effective cross-modal transfer, making them useful for segmentation, classification, and other image understanding tasks under limited supervision \cite{brown2020language,touvron2023Llama,wang2023large}. As shown in Section~\ref{sec:quantitative_results}, they improve medical image segmentation by producing more stable and well-targeted attention, even in data-scarce settings.

\subsection{LLMs for Medical Image Segmentation Tasks}

LLMs are increasingly being explored in segmentation tasks, where their contextual reasoning and representational power can guide attention, refine features, and improve generalization—especially in visually complex or low-data settings. Common strategies include projecting visual features into LLMs or encoding visual tokens via bottlenecks \cite{alayrac2022flamingo,wang2024visionllm}. Some approaches integrate LLM decoders into visual pipelines \cite{wang2024adapting}, while others leverage LLMs to interpret label semantics and enrich supervision signals \cite{kumari2024leveraging}. A recent line of work embeds frozen LLM blocks into ViT encoders to enhance classification through residual attention refinement \cite{lai2024residual}. However, many of these efforts focus on image-level tasks or high-level outputs, and their potential for improving dense prediction tasks such as segmentation remains underexplored. Our method addresses this gap by embedding a frozen LLM within a ViT-based segmentation model. This enables pixel-wise improvements without prompts or explicit class-label guidance, ensuring consistent outputs across diverse input distributions.

\section{Methodology}
\label{sec:methods}

\subsection{Overall Framework Design}

\begin{figure*}
    \centering
    \includegraphics[scale=0.0675]{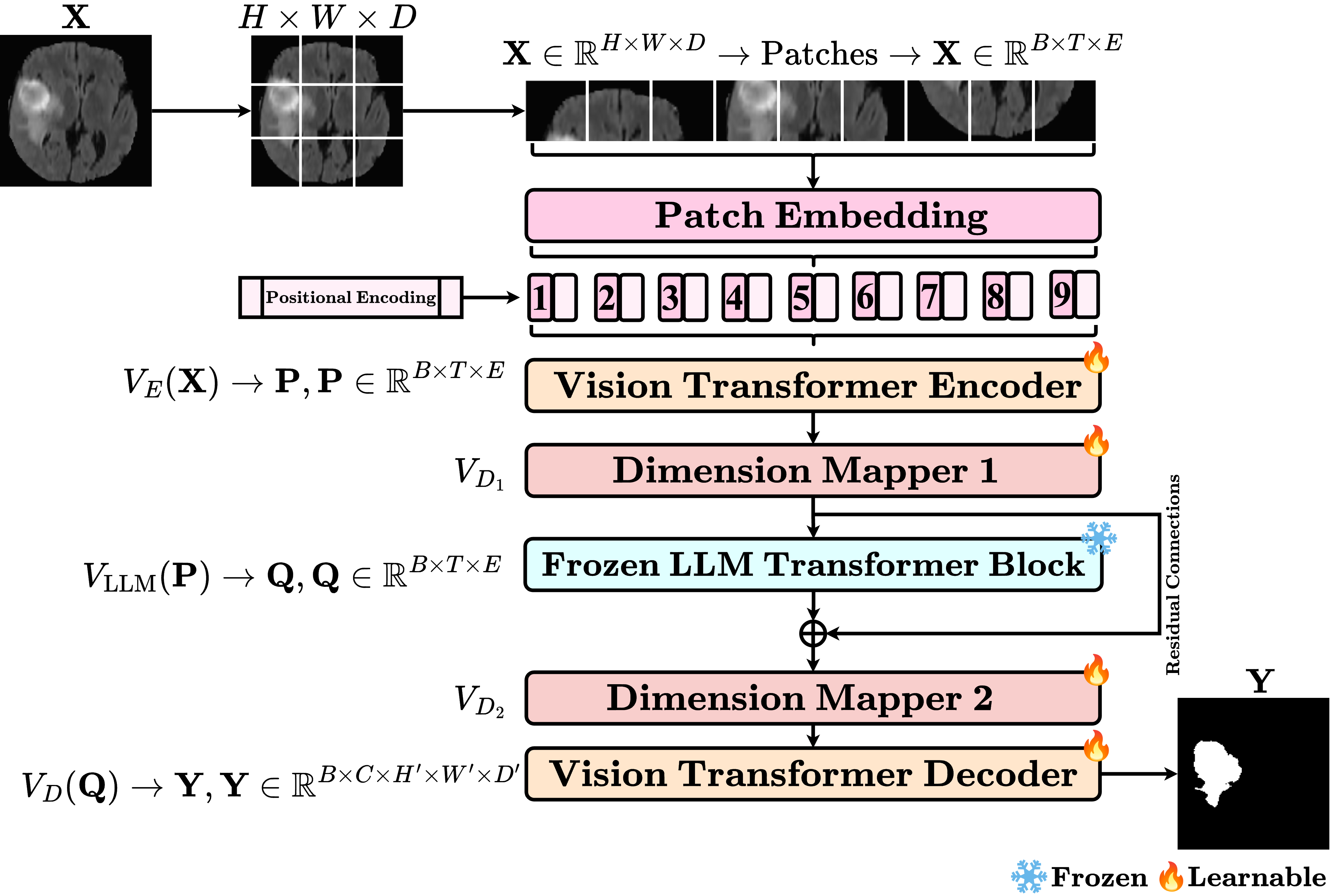}
    \caption{\textbf{Overall Framework of $\mathtt{MedVisionLlama}$ for Medical Image Segmentation.} The architecture integrates a frozen pre-trained LLM transformer block into a Vision Transformer, enhancing feature representation for 3D medical image segmentation through residual connections.}
    \label{fig:framework}
\end{figure*}

The proposed framework integrates a pre-trained Llama transformer block into a ViT-based segmentation network for 3D medical images (Figure~\ref{fig:framework}). In this study, we use Llama-3.1-8B \cite{dubey2024llama} as the pre-trained LLM, which is frozen during training to leverage its contextual knowledge and support more stable, generalizable feature representations. We selected this model for its strong semantic capacity, computational efficiency, and compatibility with reproducible, resource-constrained training environments \cite{touvron2023Llama}.

\paragraph{Base ViT Segmentation Model:}

Consider a ViT-based segmentation model where the input image \( \mathbf{X} \) is first divided into non-overlapping patches. Each patch is passed through a patch embedding module to map it into a fixed-dimensional token. Positional encodings are added to these tokens to preserve spatial structure, and the resulting sequence is fed into the ViT encoder, denoted \( V_E \). The encoder consists of hybrid attention blocks that extract both local and global visual features. This process produces a latent representation \( \mathbf{P} \), which is a sequence of enriched patch tokens. The ViT decoder, denoted \( V_D \), reconstructs the final segmentation output \( \mathbf{Y} \) from these encoded features. The overall ViT-based segmentation pipeline can be expressed as:

\begin{equation}
    V_E(\mathbf{X}) \rightarrow \mathbf{P}, \quad V_D(\mathbf{P}) \rightarrow \mathbf{Y}
    \label{eq:basic_vit}
\end{equation}

\paragraph{Enhancing Base ViT with LLM layers:}

As shown in Equation \ref{eq:basic_vit}, the standard ViT pipeline transforms the input image into a latent representation using \( V_E \), followed by decoding through \( V_D \). While effective, this structure is limited by the capacity of purely visual features to capture abstract semantic relationships. Moreover, fully leveraging this architecture often requires large amounts of labeled data, which may not be available in many medical imaging scenarios. To address this, we enhance the ViT architecture by integrating a frozen transformer block from an LLM, denoted as \( V_{\text{LLM}} \). This block leverages pre-trained knowledge obtained from large-scale text datasets to provide rich contextual features that complement visual representations, enabling better generalization in data-scarce medical imaging scenarios.

The frozen block \( V_{\text{LLM}} \) is inserted between the encoder and decoder. To bridge the dimension mismatch between the ViT encoder and the LLM block, we use two trainable LoRA-based dimension mapping layers. The first mapping layer, \( V_{D_1} \), transforms the latent visual features \( \mathbf{P} \) into the input space expected by \( V_{\text{LLM}} \). The LLM block then processes these transformed features to produce enriched embeddings, incorporating semantic context and long-range dependencies. The LLM block then projects back to the original latent space using the second LoRA-based layer \( V_{D_2} \). Its output is denoted as \( \mathbf{Q} \). This representation \( \mathbf{Q} \) is then passed to the decoder \( V_D \) to generate the final output \( \mathbf{Y} \).

\begin{equation}
    V_E(\mathbf{X}) \rightarrow \mathbf{P}, \quad V_{D_1}V_{\text{LLM}}(\mathbf{P})V_{D_2} \rightarrow \mathbf{Q}, \quad V_D(\mathbf{Q}) \rightarrow \mathbf{Y}
    \label{eq:full_vit_lora}
\end{equation}

Equation \ref{eq:full_vit_lora} represents the updated ViT architecture with integrated LLM processing and LoRA-based compatibility layers. During training, the ViT encoder \( V_E \), decoder \( V_D \), and both projection layers \( V_{D_1} \) and \( V_{D_2} \) are trainable. The LLM block \( V_{\text{LLM}} \), however, remains frozen and is only adapted through lightweight LoRA adapters. These adapters allow the network to leverage the rich knowledge encoded in the LLM without requiring full fine-tuning of the large model. Furthermore, to ensure compatibility with visual inputs, rotary positional embeddings, and attention masks originally used in the LLM's language context are removed \cite{pang2023frozen}. This hybrid architecture incorporates pre-trained language knowledge to refine features and enhance learning efficiency in vision segmentation.

\subsection{LoRA-Based Fine-Tuning for Efficient Adaptation}

We replaced conventional linear layers \cite{lai2024residual,chen2025multi} with LoRA in the dimension mapping layers ($V_{D_1}, V_{D_2}$) and applied LoRA selectively to specific layers within the Llama transformer block ($V_{\text{LLM}}$). This design enables efficient adaptation with minimal parameter updates, preserving the LLM's pre-trained knowledge and reducing computation. The mapping layers serve as lightweight adapters between ViT and LLM feature spaces. LoRA's low-rank updates allow targeted refinement of representations, improving segmentation performance with minimal overhead.
\section{Experiments}
\label{sec:experiments}

\subsection{Dataset and Preprocessing}
To evaluate our approach, we trained and tested individual models for each of the ten datasets from the Medical Segmentation Decathlon (MSD) challenge \cite{antonelli2022medical}. The tasks cover diverse anatomical structures across MRI and CT modalities. A summary of the tasks is provided in Table \ref{tab:msd_summary}. Within each dataset, we created splits for training (70\%), validation (20\%), and testing (10\%). For tasks with limited data, we applied on-the-fly data augmentation to enhance training stability and reduce overfitting.

\begin{table}[ht]
    \centering
    \scalebox{0.75}{
    \begin{tabular}{@{}lcc@{}}
        \toprule
        \textbf{Task} & \textbf{Modality} & \textbf{Number of Images} \\
        \midrule
        Task01\_BrainTumour & MRI & 484 \\
        Task02\_Heart & MRI & 20 \\
        Task03\_Liver & CT & 100 \\
        Task04\_Hippocampus & MRI & 260 \\
        Task05\_Prostate & MRI & 32 \\
        Task06\_Lung & CT & 53 \\
        Task07\_Pancreas & CT & 281 \\
        Task08\_HepaticVessel & CT & 303 \\
        Task09\_Spleen & CT & 41 \\
        Task10\_Colon & CT & 126 \\
        \bottomrule
    \end{tabular}
    }
    \caption{Summary of tasks, modalities, and number of images in the MSD dataset.}
    \label{tab:msd_summary}
\end{table}

We implemented our model in PyTorch and trained it on an NVIDIA A100 GPU (40 GB) for 200 epochs. We used a learning rate of \(2 \times 10^{-3}\), the Adam optimizer \cite{kingma2014adam}, and a batch size of 4. For segmentation, we employed a combination of Dice \cite{sudre2017generalised} and BCE loss \cite{yi2004automated}. The model was configured with an image size of \(128 \times 128 \times 128\), a patch size of \(8 \times 8 \times 8\), and a default LoRA rank of 8. For each task, model selection was performed using 5-fold cross-validation relying on the 5 non-overlapping 20\% validation sets. All final evaluations were performed on the test datasets (10\% of data) that were held out before training to ensure comparability across all comparisons.

\begin{table*}[b!]
    \centering
    \fontsize{7}{9}\selectfont
    \scalebox{0.85}{
     \begin{tabular}{@{}l@{\hskip 0.04in}|c@{\hskip 0.04in}c@{\hskip 0.04in}|c@{\hskip 0.04in}c@{\hskip 0.04in}|c@{\hskip 0.04in}c@{\hskip 0.04in}|c@{\hskip 0.04in}c@{\hskip 0.04in}|c@{\hskip 0.04in}c@{\hskip 0.04in}|c@{\hskip 0.04in}c@{\hskip 0.04in}|c@{\hskip 0.04in}c@{\hskip 0.04in}|c@{\hskip 0.04in}c@{\hskip 0.04in}|c@{\hskip 0.04in}c@{\hskip 0.04in}|c@{\hskip 0.04in}c@{}}
        \toprule
        \textbf{Metric} & \multicolumn{2}{c}{\textbf{Task01}} & \multicolumn{2}{c}{\textbf{Task02}} & \multicolumn{2}{c}{\textbf{Task03}} & \multicolumn{2}{c}{\textbf{Task04}} & \multicolumn{2}{c}{\textbf{Task05}} & \multicolumn{2}{c}{\textbf{Task06}} & \multicolumn{2}{c}{\textbf{Task07}} & \multicolumn{2}{c}{\textbf{Task08}} & \multicolumn{2}{c}{\textbf{Task09}} & \multicolumn{2}{c}{\textbf{Task10}} \\
        \midrule
        \textbf{+Llama} & \textbf{\texttimes} & \textbf{\checkmark} & \textbf{\texttimes} & \textbf{\checkmark} & \textbf{\texttimes} & \textbf{\checkmark} & \textbf{\texttimes} & \textbf{\checkmark} & \textbf{\texttimes} & \textbf{\checkmark} & \textbf{\texttimes} & \textbf{\checkmark} & \textbf{\texttimes} & \textbf{\checkmark} & \textbf{\texttimes} & \textbf{\checkmark} & \textbf{\texttimes} & \textbf{\checkmark} & \textbf{\texttimes} & \textbf{\checkmark} \\
        \midrule
        \textbf{Dice} & 0.74 & \textbf{0.91}$^*$ & 0.76 & \textbf{0.87} & 0.71 & \textbf{0.81}$^*$ & 0.72 & \textbf{0.84}$^*$ & 0.68 & \textbf{0.83}$^*$ & 0.76 & \textbf{0.88} & 0.81 & \textbf{0.95}$^*$ & 0.75 & \textbf{0.87}$^*$ & 0.78 & \textbf{0.90}  & 0.72 & \textbf{0.86}$^*$ \\
        & ±0.03 & \textbf{±0.04} & ±0.04 & \textbf{±0.07} & ±0.08 & \textbf{±0.02} & ±0.04 & \textbf{±0.03} & ±0.05 & \textbf{±0.06} & ±0.09 & \textbf{±0.07} & ±0.03 & \textbf{±0.04} & ±0.04 & \textbf{±0.03} & ±0.07 & \textbf{±0.08} & ±0.04 & \textbf{±0.03} \\
        \midrule
        \textbf{Jaccard} & 0.62 & \textbf{0.83}$^*$ & 0.61 & \textbf{0.73} & 0.55 & \textbf{0.68}$^*$ & 0.56 & \textbf{0.72}$^*$ & 0.52 & \textbf{0.66}  & 0.61 & \textbf{0.74}$^*$  & 0.73 & \textbf{0.90}$^*$ & 0.60 & \textbf{0.73}$^*$ & 0.64 & \textbf{0.82}  & 0.59 & \textbf{0.75}$^*$ \\
        & ±0.04 & \textbf{±0.02} & ±0.06 & \textbf{±0.10} & ±0.04 & \textbf{±0.08} & ±0.04 & \textbf{±0.04} & ±0.06 & \textbf{±0.09} & ±0.05 & \textbf{±0.08} & ±0.03 & \textbf{±0.03} & ±0.04 & \textbf{±0.04} & ±0.07 & \textbf{±0.10} & ±0.03 & \textbf{±0.04} \\
        \midrule
        \textbf{NSD} & 0.64 & \textbf{0.76}$^*$ & 0.67 & \textbf{0.74} & 0.63 & \textbf{0.71}$^*$ & 0.65 & \textbf{0.73}$^*$ & 0.61 & \textbf{0.72}  & 0.69 & \textbf{0.80}$^*$ & 0.71 & \textbf{0.86}$^*$ & 0.66 & \textbf{0.77}$^*$ & 0.68 & \textbf{0.80}  & 0.63 & \textbf{0.78}$^*$ \\
        & ±0.04 & \textbf{±0.04} & ±0.06 & \textbf{±0.09} & ±0.04 & \textbf{±0.10} & ±0.03 & \textbf{±0.03} & ±0.06 & \textbf{±0.08} & ±0.05 & \textbf{±0.07} & ±0.03 & \textbf{±0.05} & ±0.04 & \textbf{±0.05} & ±0.07 & \textbf{±0.09} & ±0.03 & \textbf{±0.04} \\
        \midrule
        \textbf{HD95} & 14.11 & \textbf{10.06}$^*$ & 14.70 & \textbf{11.50} & 15.36 & \textbf{10.34}$^*$ & 14.49 & \textbf{9.77}$^*$ & 14.85 & \textbf{10.50} & 15.58 & \textbf{11.50} & 14.33 & \textbf{9.68}$^*$ & 14.80 & \textbf{10.50}$^*$ & 14.42 & \textbf{10.06}$^*$ & 15.07 & \textbf{10.50}$^*$ \\
        & ±3.7 & \textbf{±2.2} & ±3.9 & \textbf{±2.9} & ±4.3 & \textbf{±2.3} & ±3.6 & \textbf{±1.9} & ±3.8 & \textbf{±2.6} & ±4.2 & \textbf{±2.5} & ±3.7 & \textbf{±1.9} & ±3.9 & \textbf{±2.1} & ±3.6 & \textbf{±2.8} & ±4.0 & \textbf{±2.55} \\
        \midrule
        \textbf{Specificity} & 0.93 & \textbf{0.98}$^*$ & 0.92 & \textbf{0.94} & 0.90 & \textbf{0.95}$^*$ & 0.91 & \textbf{0.96}$^*$ & 0.90 & \textbf{0.94}$^*$ & 0.92 & \textbf{0.95} & 0.95 & \textbf{0.98}$^*$ & 0.90 & \textbf{0.94} & 0.91 & \textbf{0.97} & 0.89 & \textbf{0.95} \\
        & ±0.03 & \textbf{±0.02} & ±0.07 & \textbf{±0.06} & ±0.04 & \textbf{±0.05} & ±0.03 & \textbf{±0.02} & ±0.07 & \textbf{±0.06} & ±0.06 & \textbf{±0.05} & ±0.02 & \textbf{±0.02} & ±0.04 & \textbf{±0.03} & ±0.08 & \textbf{±0.03} & ±0.04 & \textbf{±0.03} \\
        \midrule
        \textbf{Sensitivity} & 0.90 & \textbf{0.95}$^*$ & 0.88 & \textbf{0.90} & 0.87 & \textbf{0.92}$^*$ & 0.86 & \textbf{0.90}$^*$ & 0.86 & \textbf{0.92}$^*$ & 0.89 & \textbf{0.90} & 0.91 & \textbf{0.95}$^*$ & 0.88 & \textbf{0.90}$^*$ & 0.90 & \textbf{0.95}$^*$ & 0.86 & \textbf{0.93}$^*$ \\
        & ±0.04 & \textbf{±0.03} & ±0.07 & \textbf{±0.08} & ±0.04 & \textbf{±0.08} & ±0.03 & \textbf{±0.03} & ±0.08 & \textbf{±0.08} & ±0.09 & \textbf{±0.07} & ±0.04 & \textbf{±0.03} & ±0.04 & \textbf{±0.03} & ±0.09 & \textbf{±0.05} & ±0.05 & \textbf{±0.04} \\
        \bottomrule
    \end{tabular}
    }
    \caption{Quantitative comparison across 10 MSD segmentation tasks between the performance of $\mathtt{ViT\text{-}Baseline}$ (\textbf{\text{\texttimes}}) and $\mathtt{MedVisionLlama}$ (\textbf{\text{\checkmark}}). Reported metrics include Dice, Jaccard, NSD, HD95, Specificity, and Sensitivity. $\mathtt{MedVisionLlama}$ achieved better metric values than $\mathtt{ViT\text{-}Baseline}$ in all 60 comparisons, In 41 out of the 60 comparisons, $\mathtt{MedVisionLlama}$ outperformed $\mathtt{ViT\text{-}Baseline}$ in a statistically significant manner  (\( p < 0.05 \); 2-tailed paired test). Asterisks ($^*$) denote statistically significant improvements. 17 of the 19 comparisons that did not show statistically significant improvements quantified the results using the 4 tasks with the smallest number of datasets (tasks 02, 05, 06, and 09).}
    \label{tab:performance_comparison}
\end{table*}

\subsection{Baselines and comparison metrics}

Distinct from prior work, our method is developed entirely from scratch and does not rely on pre-trained visual encoders or language inputs. While existing VLMs \cite{wang2021simvlm,kim2021vilt,singh2022flava,radford2021learning} align visual and textual modalities using pre-trained components, our approach focuses solely on visual representation learning without any language supervision. To evaluate the effectiveness of the proposed $\mathtt{MedVisionLlama}$, we compare it against a standard ViT-based segmentation model, denoted as $\mathtt{ViT\text{-}Baseline}$. This baseline consists of a ViT segmentation architecture without integration of any language model components. Additionally, we include comparisons with several state-of-the-art medical image segmentation models. Quantitative evaluation is conducted using a comprehensive set of metrics: the Dice Coefficient \cite{dice1945measures} for overlap accuracy, the Jaccard Index \cite{jaccard1901etude} for similarity, and the 95th percentile of the Hausdorff Distance (HD95) \cite{huttenlocher1993comparing} for boundary precision. To further assess classification and spatial alignment performance, we also report Specificity, Sensitivity, and Normalized Surface Dice (NSD) \cite{nikolov2021clinically}.

\subsection{Quantitative Evaluation of Segmentation Performance}
\label{sec:quantitative_results}
In this section, we present a quantitative evaluation of $\mathtt{MedVisionLlama}$ compared to the $\mathtt{ViT\text{-}Baseline}$. We assess segmentation improvements resulting from the integration of pre-trained Llama weights across four dimensions: generalization across tasks (Section~\ref{sec:segmentation_comparison}), visualization of attention maps across different layers (Section~\ref{sec:feature_enhancement}), performance in low-data settings (Section~\ref{sec:low_data}), and comparison with state-of-the-art segmentation models (Section~\ref{sec:sota}). Our goal is to isolate the impact of LLM integration on core segmentation performance.

\subsubsection{Segmentation Performance Across Different Tasks}
\label{sec:segmentation_comparison}
Table~\ref{tab:performance_comparison} reports the results across 10 medical segmentation tasks, comparing $\mathtt{ViT\text{-}Baseline}$ and $\mathtt{MedVisionLlama}$. Across all tasks and evaluation metrics, $\mathtt{MedVisionLlama}$ consistently achieved higher scores. These improvements reflect the impact of integrating LLM features, particularly in scenarios with significant anatomical variability or diverse imaging modalities. The consistent gains suggest that the pre-trained Llama weights improved the ViT model's learning efficiency and segmentation accuracy. In every task, $\mathtt{MedVisionLlama}$ delivered more reliable results than the baseline, indicating enhanced generalization across domains.

\begin{figure*}[ht]
    \centering
    \includegraphics[scale=0.1]{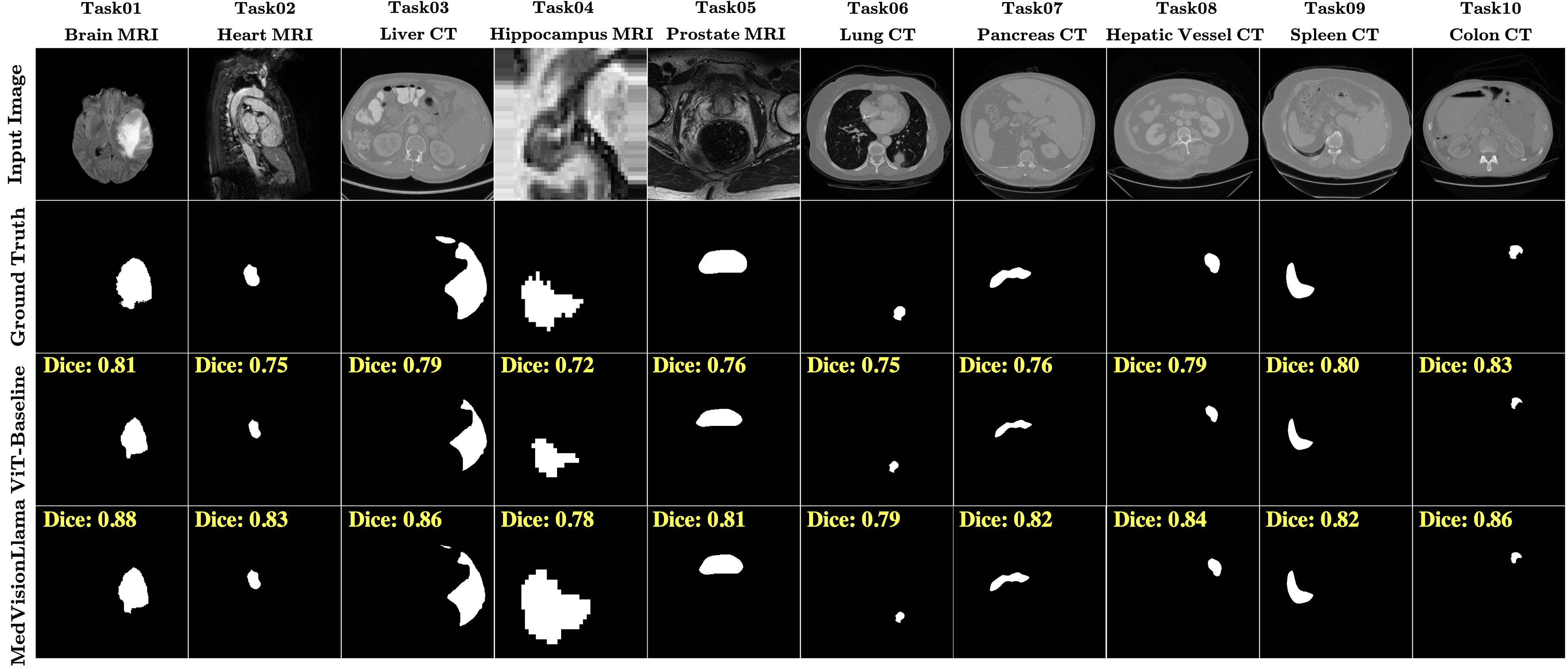}
    \caption{\textbf{Qualitative segmentation results across the 10 MSD tasks.} Top row: input images; second row: ground truth segmentations; third and fourth rows: predictions from $\mathtt{ViT\text{-}Baseline}$ and $\mathtt{MedVisionLlama}$, respectively, with corresponding Dice scores.}
    \label{fig:results_visualization}
\end{figure*}

Figure~\ref{fig:results_visualization} provides visual comparisons for representative test cases. These examples show that $\mathtt{MedVisionLlama}$ produced smoother boundaries, better object continuity, and fewer artifacts in ambiguous regions. In contrast, the baseline model often failed to accurately segment areas with unclear tissue boundaries, leading to rough edges or missing structures. The improved stability from the LLM-augmented layers contributed to more consistent and accurate predictions overall.

\subsubsection{Activation Maps: ViT-Baseline vs. MedVisionLlama}
\label{sec:feature_enhancement}
Building on the improved segmentation performance, we inspected the activation maps to understand how integrating Llama weights enhanced feature representation and helped $\mathtt{MedVisionLlama}$ focus on anatomically relevant regions. We visualized the activation maps from each layer of $\mathtt{ViT\text{-}Baseline}$ and $\mathtt{MedVisionLlama}$, comparing their ability to produce accurate and contextually meaningful representations across layers.

\begin{figure}[ht]
    \centering
    \includegraphics[scale=0.0675]{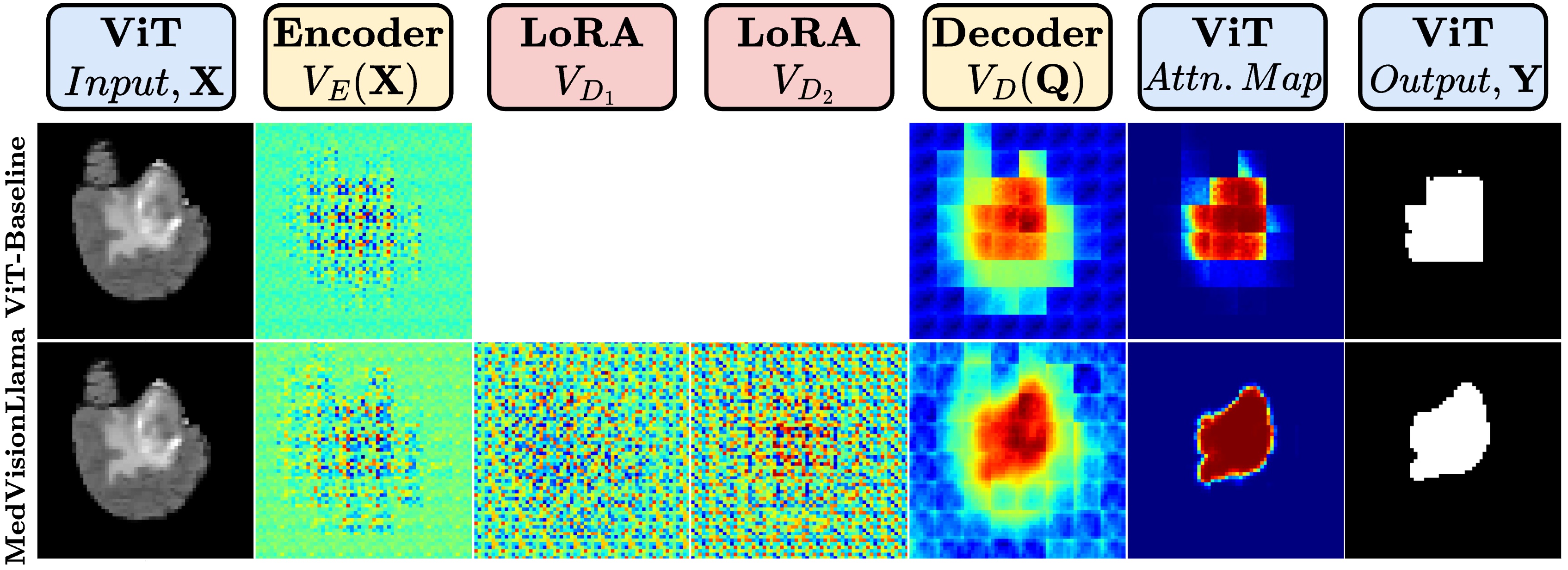}
    \caption{\textbf{Activation Maps and Attention (Task01).} Comparison of $\mathtt{ViT\text{-}Baseline}$ (top) and $\mathtt{MedVisionLlama}$ (bottom), showing activation maps for input \(\mathbf{X}\) across layers (\(V_E(\mathbf{X})\), \(V_{D_1}\), \(V_{D_2}\), \(V_D(\mathbf{Q})\)) and the final attention map leading to the output \(\mathbf{Y}\).}
    \label{fig:activation_maps}
\end{figure}

As shown in Fig.~\ref{fig:activation_maps}, Llama's additional weights improved segmentation accuracy, with LoRA dimension mapper layers refining feature extraction from the encoder's output to produce smoother predictions. In contrast, $\mathtt{ViT\text{-}Baseline}$ showed noisier activations and less precise localization. This supports our hypothesis that Llama weights, acting as residual attention boosters, enable $\mathtt{MedVisionLlama}$ to capture anatomically important regions and improve segmentation precision. These improvements were consistent across tasks, showing that LLM-based enhancements are robust and generalizable across different medical imaging types.

\subsubsection{Boosting Few-Shot Segmentation with LLM Integration}
\label{sec:low_data}
In our previous experiments, we trained $\mathtt{ViT\text{-}Baseline}$ and $\mathtt{MedVisionLlama}$ using the full dataset for each task. In this experiment, we investigated how integrating pre-trained Llama weights into the ViT architecture affected performance under data-constrained scenarios. Specifically, we evaluated both models on few-shot segmentation tasks by training with only 10\% and 30\% of the available training data for each task. Few-shot conditions mimic clinical settings with limited annotations. These experiments assess the models' ability to generalize from limited training data, which is critical for handling rare clinical cases.

\begin{figure}[ht]
    \centering
    \includegraphics[scale=0.06]{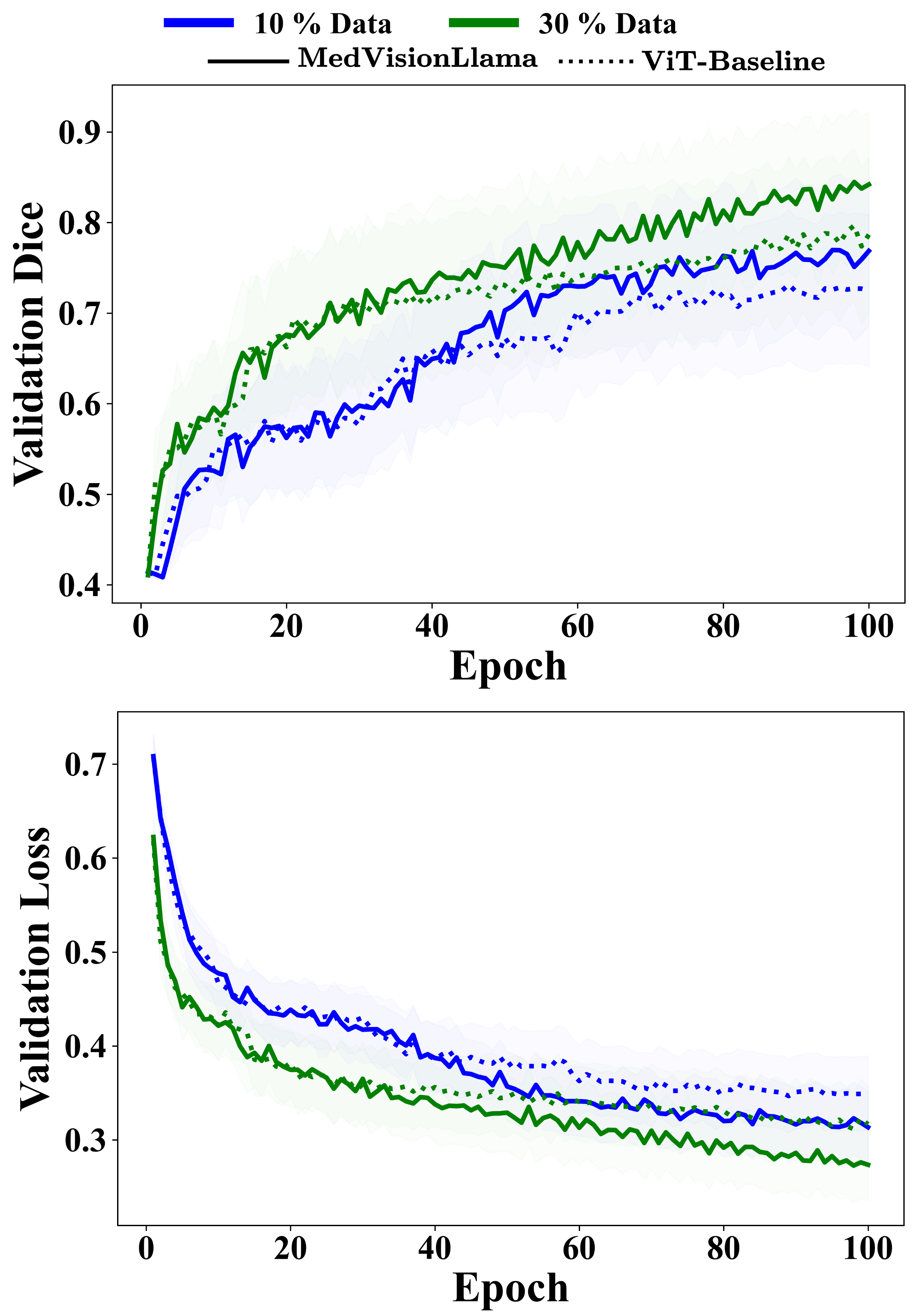}
    \caption{\textbf{Few-shot segmentation performance.} The plots show averaged validation Dice (top) and loss curves (bottom) for $\mathtt{MedVisionLlama}$ (solid lines) and $\mathtt{ViT\text{-}Baseline}$ (dotted lines), with 10\% data in blue and 30\% data in green. $\mathtt{MedVisionLlama}$ demonstrated superior performance in few-shot learning scenarios.}
    \label{fig:llm_booster}
\end{figure}

Figure~\ref{fig:llm_booster} shows that $\mathtt{MedVisionLlama}$ converged faster and outperformed $\mathtt{ViT\text{-}Baseline}$ in both validation Dice and loss curves with the 10\% data (in blue). The same trend held with the 30\% data (in green), where $\mathtt{MedVisionLlama}$ again performed well, with the enhanced feature representation resulting in better generalization averaged across all tasks. The improved few-shot performance indicated $\mathtt{MedVisionLlama}$'s ability to mitigate overfitting and enhance segmentation accuracy, reinforcing LLM integration as a data-efficient booster.

\subsubsection{Benchmarking MedVisionLlama Against State-of-the-Art Models}
\label{sec:sota}

To further validate the advantages of integrating LLM features, we benchmarked $\mathtt{MedVisionLlama}$ against several leading segmentation models on the same set of medical tasks. This comparison highlights how our approach stands in relation to established architectures, demonstrating its effectiveness not only in controlled experiments but also in broader, competitive settings.

\begin{table}[ht]
    \centering
    \scalebox{0.8}{{
    \begin{tabular}{lcc}
        \toprule
        \textbf{Model} & \textbf{Average Dice} & \textbf{Average NSD} \\
        \midrule
        UNet++ & 0.79 ± 0.04\textsuperscript{*} & 0.70 ± 0.05\textsuperscript{*} \\
        UNETR & 0.77 ± 0.05\textsuperscript{*} & 0.68 ± 0.06\textsuperscript{*} \\
        nnU-Net & 0.81 ± 0.05 & 0.71 ± 0.05 \\
        MissFormer & 0.84 ± 0.04 & 0.72 ± 0.03 \\
        TransUNet & 0.82 ± 0.06 & 0.69 ± 0.04\textsuperscript{*} \\
        Swin-UNet & 0.85 ± 0.05 & 0.74 ± 0.05 \\
        $\mathtt{ViT\ Baseline}$ & 0.74 ± 0.04\textsuperscript{*} & 0.66 ± 0.03\textsuperscript{*} \\
        \textbf{$\mathtt{MedVisionLlama\ (Ours)}$} & \textbf{0.87 ± 0.04} & \textbf{0.77 ± 0.05} \\
        \bottomrule
    \end{tabular}
    }}
    \caption{Average and SD scores of Dice and NSD across 10 MSD tasks. $\mathtt{MedVisionLlama}$ showed improved scores relative to the scores obtained by each of the other methods in all 14 comparisons. Asterisks ($^*$) indicate comparisons in which $\mathtt{MedVisionLlama}$ showed statistically significant improvements (\( p < 0.05 \), two-tailed paired t-test).}
    \label{tab:sota_comparison}
\end{table}

As shown in Table~\ref{tab:sota_comparison}, $\mathtt{MedVisionLlama}$ obtained improved scores relative to the scores obtained by all state-of-the-art models and the $\mathtt{ViT\text{-}Baseline}$ on both Dice and NSD metrics across the evaluated tasks. $\mathtt{ViT\text{-}Baseline}$ failed to deliver promising results, potentially due to limited data and lack of semantic richness, which was effectively addressed by integrating pre-trained Llama weights in $\mathtt{MedVisionLlama}$. While models such as Swin-UNet and MissFormer performed well, they did not match the consistent gains enabled by LLM integration. These results highlight the ability of $\mathtt{MedVisionLlama}$ to generalize across diverse medical imaging datasets, demonstrating the value of incorporating language-model-derived representations into visual segmentation frameworks.

\subsection{Ablation Studies}
\label{sec:ablation_studies}
With the proposed model showing promising results, we conducted a series of ablation studies to better understand the factors driving its improved performance. These investigations focused on comparing deeper ViT variants (Section \ref{sec:vit_vs_llm}), evaluating domain-specific LLMs (Table \ref{tab:evaluate_med_llm}), and assessing LoRA adaptation against linear alignment for fine-tuning Llama layers (Table \ref{tab:optimize_lora}). The goal was to isolate key contributors to the gains and determine the impact of architectural choices and complexity.

\subsubsection{MedVisionLlama: Performance Gains or Added Complexity?}
\label{sec:vit_vs_llm}

We introduced two $\mathtt{ViT\text{-}Baseline}$ variants with parameter counts comparable to $\mathtt{MedVisionLlama}$ to evaluate whether performance gains stemmed from added parameters or pre-trained Llama weights: (1) $\mathtt{ViT-Depth}$, with increased embedding size, transformer blocks, and attention heads, and (2) $\mathtt{ViT-MLP}$, with a large multilayer perceptron (MLP) incorporated into the original $\mathtt{ViT\text{-}Baseline}$. Both models were designed to closely match $\mathtt{MedVisionLlama}$ in size, ensuring that any observed improvements could be attributed to LLM-based enhancements rather than sheer model's number of parameters.

\begin{table*}[ht]
\centering
\scalebox{0.6}{
\begin{tabular}{lccccccccccccc}
\toprule
& \textbf{Number of} & & \textbf{Inference} & \multicolumn{10}{c}{\textbf{Dice Score}} \\
& \textbf{Parameters} & \textbf{GFLOPs} & \textbf{Time} & & & & & & & & & & \\
\textbf{Model} & \textbf{(in millions)} & \textbf{(per sample)} & \textbf{(ms/sample)} & \textbf{Task01} & \textbf{Task02} & \textbf{Task03} & \textbf{Task04} & \textbf{Task05} & \textbf{Task06} & \textbf{Task07} & \textbf{Task08} & \textbf{Task09} & \textbf{Task10} \\
\midrule
$\mathtt{ViT\text{-}MLP}$ & 220.45 & 0.43 & 5.75 & 0.85 & 0.74 & 0.82 & 0.77 & 0.74 & 0.82 & 0.85 & 0.79 & 0.83 & 0.85 \\
        &        &      &      & $\pm$0.04 & $\pm$0.06 & $\pm$0.07 & $\pm$0.06 & $\pm$0.04 & $\pm$0.05 & $\pm$0.06 & $\pm$0.04 & $\pm$0.03 & $\pm$0.06 \\
\midrule
$\mathtt{ViT\text{-}Depth}$ & 223.24 & 0.48 & 5.86 & 0.86 & 0.76 & 0.80 & 0.74 & 0.75 & 0.83 & 0.84 & 0.83 & 0.84 & 0.81 \\
          &        &      &      & $\pm$0.06 & $\pm$0.05 & $\pm$0.04 & $\pm$0.03 & $\pm$0.07 & $\pm$0.07 & $\pm$0.06 & $\pm$0.04 & $\pm$0.05 & $\pm$0.03 \\
\midrule
\textbf{$\mathtt{MedVisionLlama\ (Ours)}$} & \textbf{223.67} & \textbf{0.45} & \textbf{6.48} & \textbf{0.91} & \textbf{0.87} & \textbf{0.81} & \textbf{0.84} & \textbf{0.83} & \textbf{0.88} & \textbf{0.95} & \textbf{0.87} & \textbf{0.90} & \textbf{0.86} \\
                        &                &               &               & $\pm$\textbf{0.03} & $\pm$\textbf{0.04} & $\pm$\textbf{0.04} & $\pm$\textbf{0.05} & $\pm$\textbf{0.04} & $\pm$\textbf{0.05} & $\pm$\textbf{0.04} & $\pm$\textbf{0.05} & $\pm$\textbf{0.04} & $\pm$\textbf{0.03} \\
\bottomrule
\end{tabular}
}
\caption{Comparison of $\mathtt{ViT\text{-}MLP}$, $\mathtt{ViT\text{-}Depth}$, and $\mathtt{MedVisionLlama}$ on 10 MSD segmentation tasks. $\mathtt{MedVisionLlama}$ consistently outperforms both variants.}
\label{tab:vit_vs_llm}
\end{table*}

As shown in Table~\ref{tab:vit_vs_llm}, $\mathtt{MedVisionLlama}$ outperformed both variants across all tasks, suggesting that the gains stemmed from Llama's pre-trained weights. These weights acted as residual attention boosters, refining feature extraction and improving focus on anatomical structures, rather than merely increasing depth or parameters. Even in tasks where deeper ViTs showed marginal improvement, they lacked the consistency observed in LLM-enhanced representations. This result reinforces the benefit of integrating cross-modal pretraining instead of relying solely on increasing model depth or parameters.

\subsubsection{Evaluating Medical LLMs: Can They Outperform Llama?}
\label{tab:evaluate_med_llm}

This study aimed to evaluate the impact of replacing pre-trained Llama weights with domain-specific medical LLM weights, which could potentially improve segmentation and outperform Llama due to their training on clinically relevant textual data. We compared $\mathtt{MedVisionLlama}$ with various medical LLMs, including BioGPT, ClinicalBERT, and BioBERT across all 10 tasks. Each model followed the same integration strategy, ensuring that only the underlying pre-trained weights differed. These medical LLMs have been trained on large-scale biomedical corpora and clinical notes, which theoretically should better align with the semantics of medical imaging. This study aimed to determine whether such alignment translates into improved segmentation performance.

\begin{figure}[h!]
    \centering
    \includegraphics[scale=0.0145]{Figures/boxplot.png}
    \caption{\textbf{Dice score comparison across all tasks.} Box plot illustrating the distribution of averaged Dice scores for $\mathtt{MedVisionLlama}$, $\mathtt{MedVis-BioGPT}$, $\mathtt{MedVis-BioBERT}$, and $\mathtt{MedVis-ClinicalBERT}$. Statistical analysis revealed no significant differences between the performances of models (\( p > 0.05 \)).}
    \label{fig:boxplot}
\end{figure}

From Figure~\ref{fig:boxplot}, Dice scores for $\mathtt{MedVisionLlama}$, $\mathtt{MedVision-BioGPT}$, $\mathtt{MedVision-ClinicalBERT}$, and $\mathtt{MedVision-BioBERT}$ were comparable (\( p > 0.05 \)) with no statistically significant differences. This indicates that segmentation performance relies more on the feature extraction capabilities of the heavily pre-trained LLM weights than on the domain-specific nature of the LLM pretraining. These results suggest that large-scale general language pretraining may be sufficient to effectively guide visual models for segmentation tasks. Furthermore, domain-specific language models may not provide clear advantages for vision-language alignment without additional domain adaptation or multimodal tuning.

\subsubsection{Optimizing MedVisionLlama: Is LoRA the Key?}
\label{tab:optimize_lora}

\paragraph{LoRA vs. Linear Projections: }
We evaluated whether integrating LoRA into $\mathtt{MedVisionLlama}$ improves performance over using a frozen LLM with linear projection layers \cite{lai2024residual,chen2025multi}. The comparison focuses on segmentation accuracy, parameter distribution, and inference time to assess how effectively LoRA adapts the LLM for this task.

\begin{table}[ht]
\centering
\scalebox{0.56}{
\begin{tabular}{lccccc}
\toprule
\textbf{Model} & \multicolumn{2}{c}{\textbf{Parameters (in millions)}} & \textbf{Average Dice} & \textbf{Inference Time} \\
& \textbf{Non-trainable} & \textbf{Trainable} & & \textbf{(ms/sample)} \\
\midrule
\large{$\mathtt{ViT\text{-}Baseline}$} & -- & \large{1.16} & \large{0.74} $\pm$ \large{0.04} & \large{2.47} \\
\large{\textbf{$\mathtt{MedVisionLlama\ (LoRA)}$}} & \large{\textbf{218.24}} & \large{\textbf{5.43}} & \large{\textbf{0.87} $\pm$ \textbf{0.04}} & \large{6.48} \\
\large{$\mathtt{MedVisionLlama\ (Linear)}$} & \large{218.12} & \large{6.78} & \large{0.81} $\pm$ \large{0.06} & \large{7.32} \\
\bottomrule
\end{tabular}}
\caption{Comparison of $\mathtt{ViT\text{-}Baseline}$, $\mathtt{MedVisionLlama}$ with LoRA adaptation, and $\mathtt{MedVisionLlama}$ using linear projections. LoRA provides the best performance with fewer trainable parameters and moderate computational overhead.}
\label{tab:lora_vs_frozen}
\end{table}

As shown in Table~\ref{tab:lora_vs_frozen}, the number of trainable parameters in $\mathtt{MedVisionLlama\ (LoRA)}$ is only slightly higher than in the lightweight $\mathtt{ViT\text{-}Baseline}$, with a modest increase in inference time. Despite this small overhead, the LoRA-enhanced model achieves a substantial improvement in segmentation accuracy. A minimal increase in training cost yields significantly better performance, making it a practical tradeoff for accuracy-critical applications. The linear variant, using larger projection layers without fine-tuning internal weights, performs worse despite more parameters and slower inference. Both variants use the frozen Llama backbone, relying only on the final transformer block to reduce computation. Overall, LoRA adapts large LLMs more efficiently and effectively than linear projections.

\paragraph{Effect of LoRA Rank: }  
Next, we investigated how the choice of LoRA rank affects segmentation performance and parameter efficiency. We evaluated LoRA ranks 2, 4, 8, and 16 across all segmentation tasks using consistent training protocols and metrics. This analysis helps identify the optimal trade-off between model complexity and accuracy in adapting the frozen LLM.

\begin{table}[ht]
\centering
\scalebox{0.725}{
\begin{tabular}{cccc}
\toprule
\textbf{LoRA Rank} & \textbf{Trainable Params. (M)} & \textbf{Average Dice} & \textbf{Average NSD} \\
\midrule
2 & 5.04 & 0.85 $\pm$ 0.05 & 0.75 $\pm$ 0.06 \\
\textbf{4} & \textbf{5.23} & \textbf{0.89 $\pm$ 0.05} & \textbf{0.78 $\pm$ 0.06} \\
8 & 5.43 & 0.87 $\pm$ 0.04 & 0.77 $\pm$ 0.05 \\
16 & 5.62 & 0.85 $\pm$ 0.06 & 0.74 $\pm$ 0.04 \\
\bottomrule
\end{tabular}}
\caption{Ablation study on the effect of LoRA rank in $\mathtt{MedVisionLlama}$. Rank 4 achieved the best trade-off between segmentation accuracy and parameter efficiency.}
\label{tab:lora_rank}
\end{table}

Table~\ref{tab:lora_rank} shows that LoRA rank 4 yielded the best overall performance, achieving the highest average Dice and NSD with only 5.23M trainable parameters. LoRA rank determines adapter capacity and how well the frozen LLM adapts. Low ranks (e.g., 2) lack capacity, while high ranks (8 or 16) add parameters without accuracy gains and may slightly degrade performance. Moderate ranks provide the best balance, making LoRA practical for resource-aware medical segmentation.
\section{Discussion and Conclusion}
\label{sec:discussion_conclusion}

Segmentation in medical imaging continues to be challenging when data is scarce or diverse, often requiring models that can generalize well while being efficient to train. In this work, we explored whether LLMs, which are typically trained on text, could be repurposed to support visual tasks by serving as attention boosters within vision transformers. This led us to design $\mathtt{MedVisionLlama}$, a hybrid architecture that embeds frozen LLM blocks within a standard ViT backbone using a residual attention pathway. From the outset, our hypothesis was simple: LLMs, especially ones as expressive as Llama, encode rich relational structures that could enhance the attention dynamics in vision models. Instead of training these massive models from scratch, we kept them frozen and used LoRA to adapt a small set of attention weights. This let us reuse its pre-trained features with minimal computation and stable training.

The quantitative evaluations confirmed our hypothesis. Across multiple datasets and segmentation tasks, $\mathtt{MedVisionLlama}$ showed clear improvements over the baseline ViT model. It not only scored higher on standard metrics such as Dice and Jaccard but also showed better boundary delineation (lower HD95), improved sensitivity, and stronger consistency across cases, even when trained with limited supervision. Activation maps revealed enhanced focus on relevant anatomical regions, indicating more effective feature extraction. These improvements translate into more reliable and accurate segmentations critical for clinical decision-making. Additionally, the proposed approach performed competitively against several state-of-the-art methods, underscoring its robustness and generalizability across diverse medical imaging tasks.

To better understand the origin of these gains, we performed several ablation studies. Replacing Llama layers with deeper ViT blocks or MLP-based variants did not replicate the same improvements. Similarly, swapping in domain-specific LLMs such as BioGPT and ClinicalBERT yielded no consistent benefit, suggesting that the representational strength of LLMs, even without domain-specific pretraining, plays a more decisive role than previously expected. This also indicates that LLMs, even trained purely on text, carry abstract structural priors that are transferrable to visual tasks. We also examined the role of LoRA adaptation. Selectively tuning a small subset of attention weights proved effective, achieving strong performance with low training cost and memory usage—suitable for clinical or resource-constrained environments. Rank ablation further revealed that a moderate LoRA rank (e.g., 4) offers the best trade-off between accuracy and parameter efficiency.

In essence, $\mathtt{MedVisionLlama}$ introduces a simple but effective idea: using the representational structure of a frozen LLM to guide attention in a vision transformer. This approach improves performance, speeds up convergence, and keeps training cost low. Rather than relying on brute force scaling or extensive domain-specific tuning, it takes advantage of what general-purpose LLMs already know, offering a new way to build stronger segmentation models from existing components.

\paragraph{Acknowledgements: } This project was supported by funds from U.S. Deparrment of Defense CDMRP Vision Research Program Grant \# W81XWH1910853 awarded to Drs. Leonard Levin, Janine Mendola and Amir Shmuel.

{
    \small
    \bibliographystyle{ieeenat_fullname}
    \bibliography{egbib.bib}
}

\end{document}